\documentclass[12pt]{article} %
\usepackage{fleqn,amssymb,amsmath,amscd,epsfig,version,theorem}

\newcommand{\Av}{{\rm A\!v}}

\newcommand {\eh}{{\textstyle \frac{1}{2}}}
\newcommand{\beq}{\begin{equation}}
\newcommand{\eeq}{\end{equation}}
\newcommand{\Leq}[1]{\label{#1}\end{equation}}
\newcommand{\beqn}{\begin{eqnarray}}
\newcommand{\eeqn}{\end{eqnarray}}
\newcommand{\beqno}{\begin{eqnarray*}}
\newcommand{\eeqno}{\end{eqnarray*}}
\newcommand{\bem}{\le(\! \begin{array}}
\newcommand{\eem}{\end{array}\!\ri)}
\newcommand{\bsm}{\left(\begin{smallmatrix}} 
\newcommand{\esm}{\end{smallmatrix}\right)}  
\renewcommand {\le}{\left}
\newcommand {\ri}{\right}
\newcommand {\llll}{\lambda}

\def\be{\begin{equation}}
\def\ee{\end{equation}}
\def\bea{\begin{eqnarray}}
\def\eea{\end{eqnarray}}

\def\l   {\lambda}

\def\bC {{\mathbb C}}
\def\bE {{\mathbb E}}
\def\bN {{\mathbb N}}
\def\bR {{\mathbb R}}
\def\bZ {{\mathbb Z}}

\newcounter{masectionnumber}
\newcommand{\masect}[1]{\setcounter{equation}{0}
  \refstepcounter{masectionnumber} \vspace{1truecm plus 1cm} \noindent
    {\large\bf \arabic{masectionnumber}. #1}\par \vspace{.2cm}
      \addcontentsline{toc}{section}{\arabic{masectionnumber}. #1}
    }

    \newcounter{masubsectionnumber}[masectionnumber]
\newtheorem{lem}{Lemma}
\newtheorem{thm}[lem]{Theorem}
\newtheorem{prop}[lem]{Proposition}

\newtheorem{df}[lem]{Definition}

\theorembodyfont{\normalfont}
\newtheorem {example}[lem]{Example}

\newtheorem {remark}[lem]{Remark}

\newcommand {\qmbox}[1]{\quad\mbox{#1}\quad}
\newcommand {\ar}{\rightarrow}
\newcommand {\LA}{\left\langle}
\newcommand {\RA}{\right\rangle}
\begin{document}
\title{Stochastically Stable Quenched Measures}
\date{Revised, July 14, 2004}
\maketitle
\centerline{
{Alessandra Bianchi\footnote{bianchi@mat.uniroma3.it, Dipartimento di Matematica,
Universit\`a di Roma 3, Roma, Italy}, 
Pierluigi Contucci\footnote{contucci@dm.unibo.it,
Dipartimento di Matematica, Universit\`a di Bologna, 40127 Bologna, Italy}, 
Andreas Knauf\footnote{knauf@mi.uni-erlangen.de, Mathematisches Institut,
University Erlangen-Nuremberg. Bismarckstr. 1~1/2, 91054 Erlangen, Germany. 
}}}
\vskip 2truecm
\centerline{\bf Abstract}
We analyze a class of stochastically stable quenched measures. 
We prove that stochastic stability is fully characterized by an infinite family
of zero average polynomials in the covariance matrix entries.
\newpage
\masect{Introduction} \vspace{-.6cm}
\vskip .5truecm

After three decades from their first appearance in the Edwards and Anderson work 
\cite{EA} the spin glass models
and their low temperature phase remain one of the major unsolved
problems of condensed matter physics. 
The physically well understood mean field case (see Parisi et al. \cite{MPV}) 
is still under investigation from the mathematically rigorous perspective 
and some recent results by Guerra and by Talagrand \cite{G1,T2} confirm the Parisi theory. 
The case of spin glasses on finite dimensional lattices is 
instead much more controversial, and the structure of its equilibrium states
is an unsettled matter even from the theoretical physics point of view.

The spin glass model problems arise from a peculiar mathematical structure
of two intertwined probability measures, the configurational (spins) 
and the disordered (random couplings) which are combined in a precise
measure prescription of equilibrium statistical mechanics commonly called 
{\it quenched ensemble}. 

In Aizenman and Contucci \cite{AC} a stability property for the mean field models was 
derived from the continuity (in the temperature) of the 
thermodynamic functions. Some of the consequences of such 
a stability were proved to be captured by an infinite family of 
zero average polynomials (see also Ghirlanda and Guerra \cite{GG}). 
Subsequently in \cite{C} the same property
was investigated in finite dimensional models and proved to imply a formally
similar property in terms not of the standard overlap function but 
of the so called link-overlap.

The stability property, nowadays called {\it stochastic stability}, attracted
some attention from both theoretical and mathematical physics. It was
first investigated in Franz et al. \cite{FMPP1,FMPP2} and cleverly used to determine a 
relation between the off-equilibrium dynamics and the static properties.
More recently a purely probabilistic version of it expressed in terms
of invariance under reshuffling of random measure for points in the real line
has been investigated and completely classified in the case
of independent jump distribution (see Ruzmaikina and Aizenman \cite{RA}). 
The connection between the probabilistic
approach and the statistical mechanics one is well explained in \cite{G2}
and based on a new variational principle introduced in \cite{ASS}.

Can we give a complete characterization of the stochastic stability property
within its original statistical mechanics formulation? In other terms once
we know that a spin glass model verifies stochastic stability do we know
what are (all) the constraints of its overlap distribution? In this paper
we answer positively the previous questions and we prove that thanks to a 
remarkable cancellation mechanics already observed
in \cite{C2}, the zero overlap polynomials of \cite{AC} or \cite{C} provide 
a complete description of the mentioned stability property.

The paper is organized as follows: in Section 2 the quenched measure is introduced
and the overlap moments formalism explained. Section 3 introduces a combinatorial 
description of the overlap measure on graph theoretical grounds. Section 4 
introduces stochastic stability and contains the main result (Theorem \ref{thm:stabth}). 
It states a property which implies that all the consequences of the stochastic 
stability are indeed contained in 
its second order version.

\masect{Quenched measures}\label{sect:qm}
A {\em quenched probability space} is a product measurable space 
$\Omega_J\times\Omega_\sigma$, where the random probability measure 
$\mu_J$ on $\Omega_\sigma$ is indexed by $J\in \Omega_J$, and 
distributed according to a probability measure $\nu$ on $\Omega_J$.

\begin{example}\label{ex:1}
The {\em  Sherrington-Kirkpatrick} (SK) model with $N$ spins (no external field). 
\\ 
Here $\Omega_\sigma:=\{-1,1\}^N$, 
$\Omega_J:= \bR^{N^2}$, random interactions 
$J_{i,j}$ with Cartesian coordinates $1\leq i,j\leq N$
and Gaussian measure $\nu$ of density 
$(2\pi)^{-{N^2}/2}\exp(-\sum_{1\leq i,j\leq N}J^2_{i,j}/2)$ 
w.r.t.\ Lebesgue measure.
For inverse temperature $\beta\geq0$ the random
Gibbs measure $J\mapsto \mu_J$ is given by
\be
\mu_J(\sigma):=\frac{\exp(-\beta H_J(\sigma))}{Z_J(\beta)}
\qquad 
\big(\sigma \equiv (\sigma(1),\ldots,\sigma(N))\in \Omega_\sigma\big),
\label{czz}
\ee
\be
H_J(\sigma):=-\sum_{1\leq i,j \leq N}J_{i,j}\sigma(i)\sigma(j) \; ,
\eeq
and partition function 
$Z_J(\beta):=\sum_{\sigma\in \Omega_\sigma}\exp(-\beta H_J(\sigma))$.
\end{example}
\begin{example}\label{ex:EA}
The {\em  Edwards-Anderson} (EA) model with $N$ spins in a volume $\Lambda\subset \bZ^d$
and nearest neighbor interactions.\\ 
Indicating by $B(\Lambda)$ the set of nearest neighbors of $\Lambda$: 
$\Omega_\sigma:=\{-1,1\}^{|\Lambda|}$, $\Omega_J:= \bR^{|B(\Lambda)|}$, random interactions
$J_{(i,j)}$ for nearest neighbors in Cartesian coordinates $(i,j)\in B(\Lambda)$ 
and Gaussian measure $\nu$ of density
$(2\pi)^{-{|B(\Lambda)|}/2}\exp(-\sum_{(i,j)\in B(\Lambda)}J^2_{i,j}/2)$ 
w.r.t.\ Lebesgue measure. As in the former example the random Gibbs measure is given by
(\ref{czz}) with the Hamiltonian
\be
H_J(\sigma):=-\sum_{(i,j)\in B(\Lambda)}J_{i,j}\sigma(i)\sigma(j) \; .
\Leq{ea}
\end{example}
For random variables $f:\Omega_J\times\Omega_\sigma\ar\bR$ 
resp.\  $g:\Omega_J\ar\bR$ we use the notation
\[\LA f\RA (J) \ := \ \int_{\Omega_\sigma} f(\sigma,J)\, d\mu_J(\sigma) 
\qmbox{,}
\Av \left( g \right) \ := \ \int_{\Omega_J}g (J)\, d\nu(J)\ .\]
Quantities of particular interest in a quenched probability space are
the moments
$\Av \left( \LA f\RA^r \right)$ $(r\in\bN)$.
We denote the elements of the product space $\Omega_\sigma^R$ of $R\in\bN$ 
{\em real replica} by $\sigma\equiv(\sigma_1,\ldots,\sigma_R)$, and
equip it 
with product random measure 
$J\mapsto \mu_J^R\equiv \mu_J\otimes\ldots\otimes\mu_J$.
Expectation w.r.t.\ $\mu_J$ is denoted by $\ll -  \gg(J)$.\\  
The (non-random) {\em quenched measure} on $\Omega_J\times\Omega_\sigma^R$
has expectation
\[\bE( - ) \ := \ \Av\left(\ll - \gg \right).\]
It is not necessary to specify the number $R$ of replica, as a function on
$\Omega_\sigma^R$ can be extended (by inverse projection) to $\Omega_\sigma^{R+1}$.

Our plan is to study a specific class of quenched models whose algebra
of observables is built over families of Gaussian variables. Later on 
we additionally assume that the two measures involved fulfill a 
remarkable stability property.

By enlarging $(\Omega_J,\nu)$ we introduce additional 
standard normal random variables
\beq
h^{(k)}(\sigma):\Omega_J\ar\bR
\qquad \mbox{indexed by }\sigma\in\Omega_\sigma\mbox{ and } k=1,\ldots,K.
\Leq{sgrv}
We assume that the $h^{(k)}(\sigma)$ are independent of the former $J$ variables,
that $h^{(k)}(\sigma)$ is independent from $h^{(k')}(\sigma')$ if $k\neq k'$ and 
and that their covariances
\[c_{\sigma,\sigma'}:=\Av\big(h^{(k)}(\sigma)h^{(k)}(\sigma')\big)\qquad 
(\sigma,\sigma'\in\Omega_\sigma)\]
do not depend on $k\in\{1,\ldots,K\}$.
\begin{example}
For the SK model of Example \ref{ex:1} 
and independent standard normal variables 
$J^{(k)}_{i,j}$ ($i,j=1,\ldots,N; k=1,\ldots,K$)
the Gaussian variables 
\[h^{(k)}(\sigma):=N^{-1}\sum_{i,j=1}^N J^{(k)}_{i,j}\sigma(i)\sigma(j) \qquad 
(\sigma \equiv (\sigma(1),\ldots,\sigma(N))\in\Omega_\sigma)\]
have covariances 
$\Av\big(h^{(k)}(\sigma)h^{(k')}(\sigma')\big) = 
\delta_{k,k'}c_{\sigma,\sigma'}$ with $2^N\times 2^N$ matrix entries explicitly given by
\[c_{\sigma,\sigma'} := \left( N^{-1} \sum_{i=1}^N \sigma(i) \sigma'(i) \right)^2 \in [0,1]. \]
\end{example}
\begin{example}
Equivalently for the EA model of Example \ref{ex:EA} 
the Gaussian variables
\[h^{(k)}(\sigma):= |B(\Lambda)|^{-1/2}\sum_{(i,j)\in B(\Lambda)
} J^{(k)}_{i,j}\sigma(i)\sigma(j) \qquad 
(\sigma \equiv (\sigma(1),\ldots,\sigma(N))\in\Omega_\sigma)\]
have covariances
\[c_{\sigma,\sigma'} :=  |B(\Lambda)|^{-1} \sum_{(i,j)\in B(\Lambda)} \sigma(i)\sigma(j) 
\sigma'(i) \sigma'(j) \in [-1,1]. \]
\end{example}

Finally, for the replica space $\Omega_\sigma^R$ we introduce
\[c_{r,r'}: \Omega_\sigma^R\ar[-1,1]\qmbox{,}
c_{r,r'}\le(\sigma_1,\ldots,\sigma_R\ri):= 
c_{\sigma_r,\sigma_{r'}} \qquad(r,r'=1,\ldots,R).\]
The indices $r$ enumerate the replica.
By the normality assumption
\[c_{r,r}(\sigma)=1\qquad(r=1,\ldots,R;\,\sigma\in \Omega_\sigma^R).\]

Using Wick's Theorem for a 
family of Gaussian random variables, see Glimm and Jaffe \cite{GJ} 
or Simon \cite{S},
expectations of products of $h^{(k)}(\sigma)$ lead to sums of products
of $c_{r,r'}$: Whereas averages over odd products vanish, for $R=2m$
\beqno\Av\le( \prod_{i=1}^{R}h^{(k)}(\sigma_i)\ri)  & = &
\sum_{{\rm pairings}\ \pi} \prod_{i=1}^m
\Av\le(h^{(k)}(\sigma_{\pi(2i-1)}),h^{(k)}(\sigma_{\pi(2i)})\ri)\\
&=&\sum_{{\rm pairings}\ \pi} \prod_{i=1}^m
c_{\sigma_{\pi(2i-1)},\sigma_{\pi(2i)}}
=
\sum_{{\rm pairings}\ \pi} \prod_{i=1}^m
c_{{\pi(2i-1)},{\pi(2i)}}(\sigma)\,.\eeqno
Here a permutation $\pi:V\ar V$ is called a {\em pairing} with ordered
pairs 
\[\big(\pi(1),\pi(2)\big),\ldots,\big(\pi(R-1),\pi(R)\big)\] 
if $\pi(2i-1)<\pi(2i)$
and $\pi(2i-1)<\pi(2i+1)$. There are $(R-1)!!$ pairings of $V$.
\begin{example}
\[ 
\Av \left(\LA h\RA^2\right)=
\int_{\Omega_J} d\nu(J)\left[ \left( \int_{\Omega_\sigma} d\mu_J(\sigma_1) h(\sigma_1,J)\right)
\left( \int_{\Omega_\sigma} d\mu_J(\sigma_2) h(\sigma_2,J)\right) \right]
\ = \ \bE(c_{1,2}),
\]
and
\beqno
\lefteqn{\Av\left(\LA h^{(1)}\RA\LA h^{(1)}h^{(2)}\RA\LA h^{(2)}\RA\right) =}\\
&\hspace*{-25mm}=& \hspace*{-15mm}\int_{\Omega_J}\! d\nu(J)\!\! \left( \int_{\Omega_\sigma^3}
d\mu_J(\sigma_1) d\mu_J(\sigma_2)d\mu_J(\sigma_3)
h^{(1)}(\sigma_1,J)h^{(1)}(\sigma_2,J)h^{(2)}(\sigma_2,J)
h^{(2)}(\sigma_3,J) \right)\\
&=& \!\!\!\! \bE(c_{1,2} \ c_{2,3}) \; .
\eeqno
\end{example}

The general task will thus be to analyze the expectations of the so-called
{\em overlap monomials}, that is, random variables on the replica space 
$\Omega_\sigma^R$ of the form
\beq
\prod_{1\leq i<j\leq R} c_{i,j}^{m_{i,j}}\qmbox{with}m_{i,j}=m_{j,i}\in\bN_0. 
\Leq{overlap:mo}
For any permutation $\pi:V\ar V$ of the replica index set $V:=\{1,\ldots,R\}$
\beq
\bE\le(\prod_{1\leq i<j\leq R} c_{i,j}^{m_{\pi(i),\pi(j)}}\ri) = 
\bE\le(\prod_{1\leq i<j\leq R} c_{i,j}^{m_{i,j}}\ri) 
\Leq{permu}
(e.g.\
$\bE(c_{1,2}c_{2,3}c_{1,4})=\bE(c_{1,2}c_{2,3}c_{3,4})$),
since the random probability measure $\mu_J^R$ is $\pi$--invariant.

The main observation of this section is that the property of the
quenched probability space can be studied as the probability
expectation $\bE$ of the random covariance matrix $C$. 

Next section introduces the suitable combinatorial language
for the study of the overlap distribution in terms of its moments.

\masect{Multigraphs and the Gaussian Operator $\delta$}
%
In order to  analyze expectations of the overlap monomials
(\ref{overlap:mo}), it is advisable to use the notion of 
{\em (edge-labeled) multigraphs}
(see e.g.\ Diestel \cite{D} for more information).

\begin{df}
A {\bf multigraph} (on $R\in \bN$ vertices)
is a finite set $E$ and a map 
\[G:E\ar [V]^2\cup V\]
with the vertex set $V=\{1,\ldots,R\}$ and 
the family $[V]^2 := \{\{i,j\} \mid i\neq j\in V\}$ 
of unordered vertex pairs.\\
We call the elements of $E':=G^{-1}(V)$ {\bf legs}, 
the elements of $E'':=G^{-1}([V]^2)= E-E'$ 
{\bf edges}.\\
%
%
%
The set of all multigraphs is denoted by ${\cal G}$, and 
\[{\cal G}=\bigcup_{m,n=0}^\infty{\cal G}^{(m,n)}\qmbox{with}
{\cal G}^{(m,n)}:= \{G\mid |E''|=m,|E'|=n\}.\]
\end{df}
We write the multiplicities $|G^{-1}(e)|$ as exponents: 
\begin{example} 
$G=\{1,2\}^{2}\{1,3\}\{2\}\in {\cal G}^{3,1}$ 
is the graph with the edge $\{1,2\}$ of multiplicity
$2$, the edge $\{1,3\}$ and the leg $\{2\}$ of multiplicity $1$, 
and $G\in {\cal G}^{(3,1)}$. 
\end{example} 
We use the shorthand ${\cal G}'':=\bigcup_m{\cal G}^{(m,0)}$ 
for the subset of multigraphs
without legs, 
and similarly  ${\cal G}':=\bigcup_n{\cal G}^{(0,n)}$
for the multigraphs without edges.

${\cal G}$ is the basis of the vector space
$\widetilde{{\cal G}} := \bC [{\cal G}]$
of finite linear combinations, and we denote subspaces spanned by
the various subsets of ${\cal G}$ by a tilde (using multiplication
of multigraphs and formal linear combinations, $\widetilde{{\cal G}}$ 
becomes an algebra, sometimes called {\em overlap algebra}).

We now introduce the linear operator 
${\cal C}\ : \ \widetilde{{\cal G}} \to \widetilde{{\cal G}}$ of 
{\it Wick contraction} (the name being justified by (\ref{IAv:CI}) below) by
\beq
{\cal C} \le(E\ri) \ := \ E'' \,{\cal C} \le( E'\ri) 
\qmbox{,}
{\cal C} \le(E'\ri) \ := 
\sum_{{\rm pairings}\ \pi}\prod_{i=1}^m\{G(\pi(2i-1)),G(\pi(2i))\} \ . 
\Leq{wick}
for the decomposition $E=E'\cup E''$ of the multigraph $G\in{\cal G}$
(here for $|E'|=2m$ the pairings
are seen as permutations $\pi:\{1,\ldots,2m\}\ar E'\cong \{1,\ldots,2m\}$).  
Note that
\[{\cal C} \le(\widetilde{{\cal G}}^{(m,2l+1)}\ri) = \{0\}\qmbox{and} 
  {\cal C} \le(\widetilde{{\cal G}}^{(m,2l)}\ri)\subseteq\widetilde{{\cal G}}^{(m+l,0)} \ .
\]
Using the bijection $I:\hat{G}\mapsto G$ between overlap monomials 
\[\hat{G}:=\le( \prod_{1\leq i<j\leq R}c_{i,j}^{m_{i,j}}\ri)
\le(\prod_{i=1}^R h_{i}^{n_i}\ri)\]
and multigraphs $G:E\ar [V]^2\cup V$ with
edge multiplicities $|G^{-1}(i,j)|=m_{i,j}$ and leg multiplicities 
$|G^{-1}(i)|=n_i$ ,
we see that 
\beq
I\Av(\hat{G})={\cal C}I(\hat{G}).
\Leq{IAv:CI}
Thus we omit the hat of $\hat{G}$ in the rest or the article.
The invariance (\ref{permu}) under permutations  
allows us to freely use the multigraph isomorphisms
induced by arbitrary relabeling of the vertices in order to calculate
expectations.

In order to prepare for the deformation of the random measure treated 
in the next section, we introduce a second linear operator:
\[
\delta \ : \ \widetilde{{\cal G}} \to \widetilde{{\cal G}}
\qmbox{,} \delta  G \ := \ \sum_{v \in V(G)} \delta_v G \ ,
\]
with
\beq
\delta_v \ := \ \delta_v^{(+)} + \delta_v^{(-)} \ ,
\qmbox{and}
\delta_v^{(+)} G \ := \  \{v\} G \qmbox{,}
\delta_v^{(-)} G \ := \ - \{v'\} G,
\Leq{truncated}
where $v'$ is the first element of $\bN$
not belonging to $V$ (so $v'=R+1$ if $V=\{1,\ldots,R\}$).
Note that
\[\delta\le( \widetilde{{\cal G}}^{(m,l)}\ri)\subseteq
\widetilde{{\cal G}}^{(m,l+1)} 
\]
and the properties 
\beq
\delta(G_1G_2)=(\delta G_1)G_2+G_1\delta G_2\qmbox{,}\delta (\emptyset)=0.
\Leq{derivation}
of a derivation ($\emptyset$ denoting the graph without legs 
and edges, the neutral element under composition of multigraphs).
\begin{example} 
\begin{itemize}
\item
$\delta_2^{(+)} \{1,2\} = \{2\}\{1,2\}$ 
\item
$\delta_3^{(-)} \{1,3\} = - \{2\}\{1,3\}$ 
\item
$\delta_3^{(-)}\delta_3^{(-)} \{1,3\} = \{4\}\{2\}\{1,3\}$.
\end{itemize}
\end{example} 

\masect{Stable Quenched Measures}\label{sect:sqm}

In this section we investigate the consequences of a stability
assumption for the couple $(\bE, C)$ for suitable random deformation. 
Stochastic stability holds, strictly speaking, for mean field and finite dimensional 
models once the thermodynamic limit is taken. While studying its consequences
we will have to consider derivatives of the moments of $(\bE, C)$ with respect to 
suitable parameters: our computations will tacitly use the fact that the thermodynamic
limit does commute with the operation of derivative as it was proved, for every order
of derivation, in the appendix of \cite{AC}.

Given a quenched probability space
with a family (\ref{sgrv}) of standard Gaussian random variables
we introduce the deformed random
measure with expectations

\beq
 \LA  - \RA_{\llll h}\ := \ \frac{\LA  - \exp{\llll h}\RA}{\LA \exp{\llll h}\RA}
 \qmbox{,} 
\ll - \gg_{\llll h} \  := \ \otimes_{r=1}^R\LA  - \RA^{(r)}_{\llll h}
  \; ,
\Leq{dinfpr}
and the corresponding deformed quenched measure:
\be
\bE_{\llll h}( - ) \ := \ \Av(\ll - \gg_{\llll h}) \; ,
\label{dmisura}
\ee
where $h=h(\sigma,J)$ is a Gaussian random
variable of covariance $c_{\sigma,\sigma'}$, $\nu$-independent from the family 
$(h^{(1)},\ldots,h^{(K)})$, say $h=h^{(K+1)}$, and $\llll\in\bR$ 
parameterizes the deviation from  $\bE=\bE_{0}$.

We denote a monomial in the covariance matrix entries by $G$. Its
expectation is always an even function of the parameter: 
\beq
\bE_{-\llll h}(G) = \bE_{\llll h}(G)\qquad (\llll\in\bR).
\Leq{even}

The quenched probability space defined in the former section is
defined to be {\em stable} when $(\bE, C)$ is isomorphic to
$(\bE_{\llll h}, C)$ for every $\llll$
or, since the entries of $C$ are bounded, when the moments of
the two measures $\bE$ and $\bE_{\llll h}$ coincide for every
$\llll$.
Stability of the measure $\bE$ under deformation means that for
every monomial $G$
\be
\bE_{\llll h}(G) \ = \ \bE (G) \; ,
\label{prop:stabmom}
\ee
and in particular, as the left hand side is $\llll$--independent: for all $G$

\[
\frac{\partial^{2n}}{\partial \llll^{2n}}
\bE_{\llll h}(G)|_{\llll=0}\ = \ 0 \;
\]
(we consider only the even derivatives because the odd ones
vanish independently of the assumption (\ref {prop:stabmom}) 
by the symmetry of the Gaussian).

Whereas typically finite spin systems are {\em not} stable, for systems 
like the SK model stability is thought of being an asymptotic property
in the thermodynamic limit $N\ar\infty$.
\begin{prop}\label{ciccia}
For every $G\in\widetilde{{\cal G}^{''}}$ and every integer $n$
\[
\frac{\partial^{2n}}{\partial \llll^{2n}} \bE_{\llll
h}(G)|_{\llll=0} = \bE({\cal C}\delta^{2n}G) ,\]
where ${\cal C}\delta^{2n}G$ is a polynomial in the matrix
entries. So the stability property (\ref{prop:stabmom}) implies 
\[\bE({\cal C}\delta^{2n}G) \ = \ 0 \]
for every $G$ and every $n$ and in particular, defining 
\[\Delta:={\cal C}\delta^{2}\]
that 
\be
\bE(\Delta G) \ = \ 0 \qquad(G\in\widetilde{{\cal G}^{''}}).\label{f} 
\ee

\end{prop}
{\bf Proof:} 
(See also Prop. 5.1 in \cite{AC}) The action of the operator $\delta$ on the multigraph basis
corresponds to the usual derivative
with respect to the parameter $\llll$: 
\[\frac{d}{d\llll} \LA  G \RA_{\llll h}\ = 
\ \LA  G h \RA_{\llll h}- \LA G \RA_{\llll h}\LA h \RA_{\llll h} \]
So such a derivative produces a {\it truncated correlation} expressed in
the rule (\ref{truncated}). \\
The derivation property (\ref{derivation}) represents the Leibniz rule
\[\frac{d}{d\llll}\ll - \gg_{\llll h} \  = 
\sum_{l=1}^R \le(\frac{d}{d\llll}\LA  - \RA^{(l)}_{\llll h}\ri)
\bigotimes_{r\neq l}\LA  - \RA^{(r)}_{\llll h}\]
for the derivative on the replica space. 
Each differentiation with respect to $\llll$ produces a sum
of monomials, each containing an additional zero mean Gaussian variable $h$.\\
Finally, the contraction ${\cal C}$ represents Wick's Theorem.\hfill $\Box$

Our main result is

\begin{thm}\label{thm:stabth}
\beq
 {\cal C}\delta^{2n} \ = \ (2n-1)!!\, \Delta^n \qquad(n\in\bN_0),
\Leq{zmonc}
or equivalently,
\be \frac{\partial^{2n}}{\partial\llll^{2n}}\bE_{\llll h}
(G)|_{\llll=0}\,=\,(2n-1)!!\,\bE(\Delta^n G)\quad(n\in\bN_0,\, G\in\widetilde{{\cal
G}^{''}}) \; .
\label{mainte}
\ee
As a consequence the stability property is equivalent to the set of
relations (\ref{f}).
\end{thm}
\begin{remark}
In the sense of formal power series  
we can combine the set of identities (\ref{zmonc}) as
\be
{\cal C}\exp(t\delta) \; = \; \exp\left(\eh
\Delta t^2\right) \; \qquad(t\in\bC),
\ee
which justifies to call $\delta$ {\em Gaussian operator}.
\end{remark}

The proof is based on the following:
\begin{lem}\label{d1}
For all $ G\in{\cal G}^{''} $, we find 
(with $\Delta={\cal C}\delta^2$)
\beq
\frac{\partial}{\partial\llll}\bE_{\llll h}(G)
\,=\,
\llll \bE_{\llll h}(\Delta G) 
\Leq{dprima}
\end{lem}
{\bf Proof:}
By definition of $\delta$, we know that for the overlap monomial the l.h.s.\
equals
\beq
\frac{\partial}{\partial\llll}\bE_{\llll h}(G)\,=\, \bE_{\l h}(\delta G) \; 
=\bE_{\l h}\le((-\,R\,h_{R+1}+\sum_{i=1}^R h_i\,)G\ri),
\Leq{left}
and  that the r.h.s.\ of (\ref{dprima}) equals
\beq
\bE_{\llll h}(\Delta G) \,
= \,\bE_{\l h}\le(\le(2\sum_{1\leq i<j\leq R}c_{i,j}-2R\sum_{i=1}^Rc_{i,R+1}+
R(R+1))c_{R+1,R+2}\ri)G\ri),
\Leq{right}
see \cite{AC}, Lemma 6.2.

We now do integration by parts in order to get rid of the
unpaired Gaussian variable that appears in the
r.h.s.\ of (\ref{left}). The same integration by parts is responsible
for the appearence of the factor $\lambda$ in $(\ref{dprima})$.
Recall that for a family $(h^{(1)},\ldots, h^{(K)})$ of Gaussian variables 
with 
covariance matrix $\{c_{l,m}\}$ the following rule of integration by parts
\beq
 \bE\Big(h^{(l)} f(h^{(1)},\ldots, h^{(K)})\Big)\,=\,
\sum_{m=1}^K\bE \le( c_{l,m}\frac{\partial}{\partial
h^{(m)}}f(h^{(1)},\ldots, h^{(K)})\ri)
\Leq{ibp}
holds (if $f\in C^1(\bR^k)$ is of moderate growth at infinity like
for $h\mapsto \exp(\llll h)/\LA\exp(\llll h)\RA$), 
see e.g. \cite{CDGG} or Talagrand \cite{T1}, A.6.

Applying (\ref{ibp}) to (\ref{left}) we obtain (\ref{right}).
 \hfill $\Box$\\
\begin{example} 
Let us explicitly verify the lemma in the easiest non-trivial
case  $G\,=\,\{1,2\}$ (see also Sec. 5 in \cite{AC} and 
Theorem 2 in \cite{G}). From (\ref{right}) we see that
$$ {\cal C}\delta ^2 \{1,2\}\,=\,2
\le[ \{1,2\}^2 - 4\{1,2\}\{2,3\}  +3\{1,2\}\{3,4\} \ri].$$
Applying to $G$ the first derivative in $\llll$ and then
integration by parts formula (\ref{ibp}), we find:
\beqno
\lefteqn{\frac{\partial}{\partial\llll}\bE_{\llll h}(c_{12})= \ 
\Av\left[ \int_{\Omega_{\sigma_1,\sigma_2}}
c_{12}(h_1+h_2)\frac{e^{\llll(h_1 +h_2)}}{Z^2}
-2\int_{\Omega_{\sigma_1,\sigma_2,\sigma_3}}
c_{12}h_3\frac{e^{\llll(h_1 +h_2 +h_3)}}{Z^3}\right]}\\
& = & 2\sum_i \Av \left[
\int_{\Omega_{\sigma_1,\sigma_2}} c_{12}c_{2i}
\frac{\partial}{\partial h_i}\left(\frac{e^{\llll(h_1+h_2)}}{Z^2}\right)-
\int_{\Omega_{\sigma_1,\sigma_2,\sigma_3}}c_{12}c_{3i}
\frac{\partial}{\partial h_i}\left(\frac{e^{\llll(h_1 +h_2+h_3)}}{Z^3}\right)\right]\\
& = & 2\llll \ \Av 
\left[\int_{\Omega_{\sigma_1,\sigma_2,\sigma_3}}c_{12}\left(
(c_{21}+1)\frac{e^{\llll(h_1 +h_2)}}{Z^2}-
2c_{23}\frac{e^{\llll(h_1+h_2 +h_3)}}{Z^3}\right)\right. \\
& & \left. - \int_{\Omega_{\sigma_1,\sigma_2,\sigma_3,\sigma_4}}
c_{12}\left((c_{31}+c_{32}+1)\frac{e^{\llll(h_1
+h_2 +h_3)}}{Z^3}-3c_{34}\frac{e^{\llll(h_1 +h_2 +h_3+h_4)}}{Z^4}\right)
\right] \\
& = & 2\llll \;\bE_{\llll h}\;[c_{12}^2 -4c_{12}c_{13}+3c_{12}c_{34}]\,.
\eeqno
The associated multigraph polynomial of this expression,
$2[\{1,2\}^2-4\{1,2\}\{2,3\}+3\{1,2\}\{3,4\}]$, is just equal to ${\cal
C}\delta^2 G$.
\end{example}
\noindent
{\bf Proof of Theorem \ref{thm:stabth}:}
Considering $G\in\widetilde{{\cal G}^{''}}$ and thanks to equality
(\ref{dprima}), we can write the next derivatives in $\llll$ as
follows:
\bea
\frac{\partial^2}{\partial \l^2} \bE_{\l h}(G) \; &=& \; \bE_{\l
h}(\Delta G) +\l \frac{\partial}{\partial \l}\bE_{\l
h}(\Delta G) \; ,
\eea
and for general $k\in\bN$ by induction
\be
\frac{\partial^k}{\partial
\l^k} \bE_{\l h}(G) \; =\; (k-1) \frac{\partial^{k-2}}{\partial
\l^{k-2}}\bE_{\l h}(\Delta G)+\l
\frac{\partial^{k-1}}{\partial \l^{k-1}}\bE_{\l h}(\Delta
G).
\ee
Considering that in $\l=0$ all the odd derivatives vanish by (\ref{even}),
we obtain:
\beq
\frac{\partial^{2n}}{\partial
\l^{2n}} \bE_{\l h}(G)|_{\l =0} \; = \; (2n-1)
\frac{\partial^{2n-2}}{\partial \l^{2n-2}}\bE_{\l h}(\Delta
G)|_{\l =0} \; .
\Leq{eq:deri}
Combining eqs.\ (\ref{eq:deri})
\[
\frac{\partial^{2n}}{\partial
\l^{2n}} \bE_{\l h}(G)|_{\l =0} \; = \; (2n-1)!!\, 
\bE(\Delta^n G).
\]
In terms of the operation $\delta$ the previous relation becomes 
\[
{\cal C}\delta^{2n} \; = \; (2n-1) {\cal C}\delta^{2n-2} \Delta \; = ... = \; (2n-1)!!\, 
\Delta^n \; ,
\]
which proves (\ref{mainte}).\\ 
To prove equivalence of (\ref{f}) with stability, 
we observe that $\Delta^n G$ is a polynomial in
the overlap algebra i.e. 
$\Delta^n G = \sum_{\alpha} n_{G_{\alpha}}G_{\alpha}$ so that the vanishing of the average
of $\Delta G$ for every $G$ implies also the vanishing of the average of $\Delta^n G$:
\be
\frac{\partial^{2n}}{\partial \llll^{2n}} \bE_{\llll}(G)|_{\llll=0} \, = \,
(2n-1)!! \bE(\Delta^n G)=
(2n-1)!!\sum_{\alpha}n_{G_{\alpha}} \bE (G_{\alpha}) \, =
\, 0
\; 
\label{prop:18}
\ee

\begin{remark}
In order to fully appreciate the content of the (\ref{zmonc}) we examine the case
$n=2$ (see also \cite{C2}):
\be
{\cal C}\delta^{4} \ = \ 3 \Delta^2 \; .
\label{zminc}
\ee
It is interesting to note that the left hand side of (\ref{zminc})
contains {\it a priori} 
$3\cdot 2^4 = 48$ terms
(the factor 3 coming from the Wick contraction of a fourth order monomial 
and $2^4$ being the number of terms
of $(\delta^+ + \delta^-)^4$ ).
The right hand side instead contains only $4^2=16$ terms coming
from the square of $\Delta$. 

Although
by definition the Wick contraction does not conserve the number of edges nor the 
number of vertices, 
the presence of alternating signs in the definition of $\delta$ together
with the  invariance under permutation of graph labellings
produces a delicate cancellation mechanism responsible 
for the clean equality (\ref{zminc}).
\end{remark}

{\bf Acknowledgments}.
One of us (PC) thanks M.Aizenman, S. Graffi and F.Guerra for many interesting
observations on the property of stochastic stability.


\vspace{-.6cm} \addcontentsline{toc}{section}{References}
\begin{thebibliography}{1}



\bibitem[AC]{AC}
M.Aizenman and P.Contucci, ``On the Stability of the Quenched State
in Mean Field Spin Glass Models,'' {\em J. Stat. Phys.}, {\bf 92}, N. 5/6,
765--783 (1998). Available at cond-mat/9712129.

\bibitem[ASS]{ASS}
M.Aizenman, R.Sims, S.Starr, ``An Extended Variational Principle for the SK Spin-Glass Model"
{\em to appear in Phys. Rev. B.} available at cond-mat/0306386



\bibitem[C]{C}
P.Contucci, ``Replica Equivalence in the Edwards-Anderson Model,''
{\em Jou. Phy. A: Math. and Gen.}, {\bf 36}, 10961--10966 (2003)

\bibitem[C2]{C2}
P.Contucci, 
``Toward a classification theorem for stochastically stable measures,''
{\em Mark. Proc. Rel. Fie}, {\bf 9}, N. 2, 167--176, (2002)

\bibitem[D]{D}
R.Diestel: ``Graph Theory,'' Graduate Texts in Mathematics, Volume 173. 
Springer 2000 

\bibitem[EA]{EA}
S.F.Edwards and P.W.Anderson. ``Theory of Spin Glasses'', {\em J.Phys.F},  
{\bf 5}, 965--974, (1975)

\bibitem[FMPP1]{FMPP1}
S.Franz, M.Mezard, G.Parisi and L.Peliti, ``Measuring equilibrium properties
in aging systems'' {\em Phys. Rev. Lett.}, {\bf 81}, 1758 (1998)

\bibitem[FMPP2]{FMPP2}
S.Franz, M.Mezard, G.Parisi and L.Peliti, ``The response of glassy systems to
random perturbations: A bridge between equilibrium and off-equilibrium''
cond-mat/9903370,  {\em J. Stat. Phys.} {\bf 97}, 459 (1999) 

\bibitem[GG]{GG} S.Ghirlanda and F.Guerra, ``General properties of 
overlap probability distributions in disordered spin systems. T
oward Parisi ultrametricity'' {\em J. Phys. A}, {\bf 31}, no. 46, 9149-9155, (1999)

\bibitem[GJ]{GJ}
J.Glimm, A.Jaffe A.:``Quantum Physics,'' Springer 1987

\bibitem[G]{G} F.Guerra, ``About the overlap distribution in a mean field
spin glass model'', {\em Int.J.Phys.B}, {\bf 10}, 1675-1684, (1997)

\bibitem[G1]{G1} F. Guerra, ``Broken Replica Symmetry Bounds in Mean Field
Spin Glass Models'', {\em Comm.Math.Phys}, {\bf 233}, 1--12 (2003)

\bibitem[G2]{G2}
F.Guerra ``About the cavity Field in Spin Glass Models", to appear in 
{\em Proceedings of ICMP Lisbon, 2003}, available at cond-mat/0307673

\bibitem[MPV]{MPV} M.M\'ezard, G.Parisi and M.Virasoro, ``Spin Glass Theory
and Beyond'' World Scientific Lecture Notes in Physics, Vol. 9, (1987)


\bibitem[RA]{RA}
A.Ruzmaikina, M.Aizenman ``Characterization of Invariant Measures
at the Leading Edge for Competing Particle Systems'' 
{\em to appear in "Annals of Probability"}.

\bibitem[S]{S}
B.Simon, ``The $P(\Phi)_2$ Euclidean (Quantum) Field Theory,'' 
Princeton University Press 1974

\bibitem[T1]{T1}
M.Talagrand, ''Spin Glasses: A Challenge for Mathematicians,''  
Ergebnisse der Mathematik und ihrer Grenzgebiete, Vol. 46, Springer 2003

\bibitem[T2]{T2} M.Talagrand, ``The Generalized Parisi Formula'', C.R.A.S., 
{\it to appear}

\bibitem[CDGG]{CDGG}
P.Contucci, M.Degli Esposti, C.Giardina and S.Graffi,
``Thermodynamical Limit for Correlated Gaussian Random Energy Models,'' 
{\em Commun. Math. Phys.},  {\bf 236},  55--63 (2003)

\end{thebibliography}
\end{document}